# Dynamic Carrier Modulation via Nonlinear Acoustoelectric Transport in van der Waals Heterostructures

Timothy J. McSorley[1], Kaustubh Simha[1], James E. Corcoran[1], Izzie J. Catanzaro[1], Haochong Zhang[1], Meitong Yin[1], Tzu-Ming Lu[2], Davis Thuillier[1], Marshall A. Campbell[1], Thomas Scaffidi[1], Luis A. Jauregui[1,*]

[1]Department of Physics and Astronomy, University of California, Irvine, CA 92697, USA

[2]Center for Integrated Nanotechnologies, Sandia National Laboratories, Albuquerque, NM 87123, USA

* Corresponding email: lajaure1@uci.edu

**Abstract**

Dynamically manipulating carriers in van der Waals heterostructures could enable solid-state quantum simulators with tunable lattice parameters. A key requirement is forming deep potential wells to reliably trap excitations. Here, we report the observation of nonlinear acoustoelectric transport and dynamic carrier modulation in boron nitride-encapsulated graphene devices coupled to intense surface acoustic waves (SAWs) on LiNbO$_3$ substrates. SAWs generate strong acoustoelectric current densities ($J_{AE}$), transitioning from linear to nonlinear regimes with increasing SAW intensity. In the nonlinear regime, periodic carrier (electrons, holes, or their mixtures) stripes emerge. Using counter-propagating SAWs, we create standing SAWs (SSAWs) to dynamically manipulate charge distributions without static gates. The saturation of $J_{AE}$, attenuation transitions, and tunable resistance peaks confirm strong carrier localization. These results establish SAWs as a powerful tool for controlling carrier dynamics in two-dimensional (2D) materials, paving the way for the development of time-dependent quantum systems and acoustic lattices for quantum simulation.

**Introduction**

Reducing the material's dimensionality to 2D or one-dimensional (1D) reveals exotic electronic phenomena driven by confinement effects. In 1D systems, interacting electrons challenge conventional Fermi liquid theory, used to describe metals and semiconductors[1–3], leading to emergent phases such as coupled spin-charge orders and electronic stripes in strongly correlated materials like copper-oxide superconductors[4–6] and nickelates[7]. These phases result from a balance between magnetic interactions, Coulomb forces favoring localization, and the kinetic energy of doped holes promoting delocalization. Similar stripe-like charge-density waves (CDWs)[8–12] and nematic phases have been observed in quantum Hall systems[13–19] and 2D material systems[20–22], providing platforms for studying 1D physics, quantum correlations, and emergent phenomena such as unconventional superconductivity.

Methods to achieve 1D correlated systems include lithographically defined wires[23], twisted bilayers[24–26], construction of heterostructures on non-Euclidian topography-patterned surfaces[27], strong magnetic fields[28,29], and quasi-2D materials[30]. An alternative approach leverages SAWs[31,32], which propagate along the surface of piezoelectric substrates and generate dynamical electric fields. SAWs are typically generated via interdigitated transducers (IDTs), which define the geometry of the waves produced along the crystal surface. SAWs have been widely applied in sensing and 2D electron gas systems (2DEGs)[33–36] and more recently in van der Waals (vdW) materials[37–43]. In graphene, SAWs bridge classical and relativistic physics, enabling acoustoelectric transport with predicted transitions between Schrödinger- and Dirac-like behavior[44]. Prior studies have explored SAW-induced currents, Landau quantization, and synthetic magnetic fields, but these have primarily involved low-intensity SAWs, limiting interaction strength[40].

Previous experiments with SAWs in monolithic GaAs/AlGaAs heterostructures have revealed essential insights into magnetoconductivity properties of 2DEGs, providing early compelling evidence for the composite fermion model of the fractional quantum Hall effect[45]. SAWs have also been used to convey single electrons through quantum point contacts[46,47]. However, GaAs has a weak electromechanical coupling coefficient ($K^2 = 0.00064$), resulting in limited carrier-SAW interaction. In contrast, substrates with higher coupling coefficients, such as 128° Y-Cut LiNbO$_3$ with $K^2 = 0.056$, enable significantly stronger coupling and more efficient carrier-SAW interactions[48]. Prior studies involving 2D materials on LiNbO$_3$ focused on using low SAW intensities[37] or measured SAW attenuation by observing changes in the SAW phase velocity[38], leaving the high-intensity acoustoelectric transport regime characterized by substantial momentum exchange between SAWs and carriers unexplored. Such nonlinear behavior has been observed in heterostructures combining LiNbO$_3$ and GaAs[49], highlighting the potential for exploring new physics in these systems.

Here, we investigate high-intensity SAW-driven acoustoelectric transport in BN-encapsulated graphene devices on LiNbO$_3$. We demonstrate the generation of significant acoustoelectric currents ($I_{AE}$) in graphene and observe a transition from linear to nonlinear acoustoelectric transport, where periodic electron or hole stripes emerge. Using counter-propagating SAWs, we create SSAWs that dynamically modulate charge distributions without relying on static gates, creating tunable barriers that affect transport in graphene. These findings establish SAWs as a powerful tool for studying 1D physics in 2D materials and pave the way for novel applications in time-dependent quantum systems and solid-state quantum simulation.

**Results**

Figure 1a shows an optical image of a BN-encapsulated graphene device fabricated on a 128° Y-cut LiNbO$_3$ substrate, positioned between two bidirectional IDTs designed to generate SAWs propagating along the $x$-axis of LiNbO$_3$. The IDTs are designed for a SAW wavelength $\lambda = 10~\mu m$. The device preparation is described under Methods in the Supporting Information. Figure 1b displays the four-probe resistance ($R_{4p}$) vs. top gate voltage ($V_{tg}$) at temperatures $T = 300~K$ and 4 K, showing the Dirac point within 0.5V, indicative of high sample quality with minimal disorder. The left inset shows a magnified image of the device and the right inset provides a schematic cross-section of the device structure.

Figure 1c displays the IDT transmission (S$_{21}$) vs. frequency ($f$), measured at $T = 300~K$ (using a Vector Network Analyzer) showing a resonant peak at $f \sim 375~MHz$, matching the design expectations. Considering the speed of sound along the $x$-axis in LiNbO$_3$ is $v_{sound} \sim 3{,}980~m/s$, the experiment is in reasonable agreement with the expected $f = v_{sound}/\lambda$. The small discrepancy can be attributed to a 150 nm deviation in the IDT dimensions from the design. Figure 1d shows the normalized longitudinal $I_{AE}$ measured in Device #1, peaking at the same $f \sim 375~MHz$ for both IDTs, consistent with the S$_{21}$ measurements. This confirms that the measured $I_{AE}$ originates from SAWs propagating on the substrate.

In previous studies on 2DEGs, the acoustoelectric current density ($J_{AE}$) was described by the semi-classical formula $J_{AE} = -\mu I \Gamma / v_{sound}$ (I), where $\mu$ is the carrier mobility, $\Gamma$ is the attenuation coefficient, and $I$ is the SAW intensity. The SAW intensity is defined as $I = 2(P/w)10^{-IL/10}$ (II), where $P$ is the RF power applied to the IDT, $w = 85~\mu m$ is the width of the IDT aperture, and $IL$ is the insertion loss of the transducers, measured to be 20 dB and 16 dB in our samples at $T = 300~K$ and 4K respectively. For low $I$, the modulation of the carrier density is minimal and the attenuation can be expressed as $\Gamma = K^2 \frac{\pi}{\lambda} \frac{\frac{\sigma}{Y\sigma_m}}{1+(\frac{\sigma}{Y\sigma_m})^2}$

(III)[50] where $\sigma$ is the sheet conductivity, $\gamma$ is the electron drift parameter ($\gamma = 1$ for no external electric field applied), and $\sigma_m$ is the characteristic conductivity which depends on the substrate parameters and is defined as $\sigma_m = v_{sound}\epsilon_0(1 + \epsilon_r)$ (IV), where $\epsilon_0$ is the vacuum permittivity, $\epsilon_r= 50$ is the relative permittivity in LiNbO$_3$[41], and $\sigma_m\sim$ 1.8x10$^{-6}$ S. In single-layer graphene, the sheet conductivity exceeds $\sigma_m$ even at the Dirac point. Therefore, $\Gamma$ in Equation III and $J_{AE}$ in Equation I can be simplified to $\Gamma = K^2 \frac{\pi}{\lambda} \frac{\sigma_m}{\sigma}$ (V) and $J_{AE} = -\frac{IK^2\pi\sigma_m}{\lambda v_{sound}}\frac{1}{en_e}$ (VI) respectively, where $e$ is the electron charge and $n_e$ is the electron density[36]. While Equation VI accurately describes single-carrier systems such as GaAs, it does not apply to two-carrier systems like graphene and other semimetals. For a two-carrier system, $J_{AE} = -\frac{IK^2\pi\sigma_m}{e\lambda v_{sound}}\frac{(n_e-n_h)}{(n_e+n_h)^2}$ (VII) where $n_h$ is the hole density[39]. In GaAs, $J_{AE}$ vs. $n_e$ peaks when $\sigma = \sigma_m$, as described by Equation III. On the other hand, in graphene, the mixed-carrier state near the charge neutral point (CNP) causes $J_{AE}$ vs. $n$ to peak when $n$ approaches a value of the order of the disorder-induced minimal carrier density, where $n = n_e - n_h$. Leading to higher $J_{AE}$ values in low-disorder graphene samples (see Supporting Information).

Figure 2a shows the measured $J_{AE}$ vs. $I$ for $I < 0.5\ W/m$ at various $n$ and $T = 300\ K$. The data exhibits a linear dependence with a slope that decreases with increasing $n$, consistent with Equation VII and previous studies on graphene and 2DEGs[37,41,49,51]. Figure 2b extends this measurement to higher $I$, up to 7 W/m, where $J_{AE}$ transitions from the linear regime to saturation. The $I$ required for saturation increases with higher $n$, a trend observed for both electrons and holes. Dashed lines represent linear fits, closely matching lower $I$ data but strongly deviating as saturation occurs.

Figure 2c displays $J_{AE}$ vs. $I$ at $T = 4\ K$, showing similar nonlinear behavior to room temperature but with some differences. $J_{AE}$ reaches higher currents than at 300 K. Using the drift velocity equation, $J_{AE} = env$ (VIII), where $v$ is the carrier velocity, we extract $v$ vs. $I$ (inset of Figure 2c), showing $v$ saturates at $v \sim 3{,}700\ m/s$ at low $n$. Beyond this, increasing $I$ no longer affects $v$, marking a transition to nonlinear transport. The saturated velocity is slightly lower than $v_{sound} = 3{,}980\ m/s$ in LiNbO$_3$ (depicted by a black dashed line in the inset). This discrepancy may arise from the disorder-induced minimal carrier density, which is predicted to reduce the ultimate carrier velocity according to the model we have developed for high intensity acoustoelectric transport in graphene (see Supporting Information). Another difference between 4K and 300K is that lower $n$ tends to revert to the linear regime, also indicating that the density dependence of the nonlinear acoustoelectric effect for graphene is strongly linked to the sharpness of the Dirac peak and thus the sample disorder. At higher $n$, $v \ll v_{sound}$, indicative of the linear regime at both temperatures.

Figure 2d shows $J_{AE}$ vs. $n$ over a broad range of $I$, revealing a peak for electrons and a dip for holes. The $n$ at which the peak or dip occurs ($n_{peak}$) remains constant at low $I$, consistent with Equations VI and VII. In two-carrier systems, $n_{peak}$ depends solely on the disorder-induced minimal carrier density, as supported by our calculations in the Supporting Information and previous studies on 2DEGs and graphene[37,39]. Our model predicts that for small SAW voltage amplitude $\Phi \sim \Delta V$, where $\Delta V$ is the disorder-induced average gate voltage, $n_{peak}$ has a constant value set by disorder, i.e. $n_{peak} \sim C\Delta V/e$, where $C$ is the capacitance. This prediction is consistent with the finite intercept of $n_{peak}$ vs. $\sqrt{I}$ shown in the inset of Fig 2d at low $\sqrt{I}$ and describes the linear regime. At high $I$, $I > (1.5)^2\ W/m$, $n_{peak}$ increases linearly with $\sqrt{I}$ as shown in the inset of Fig. 2d. We argue that $J_{AE}$ vs. $n$ peaks when the $\Phi = \sqrt{I/\alpha}$ approaches $V_{tg}$ (where $\alpha$ is a constant specific to the system). Under this condition, the modulated carrier density $n(x) = n + n'\cos(kx)$ - where $k$ is the SAW wavevector, $n'$ is the amplitude of carrier density modulation induced by SAWs, and $n$ is tuned by $V_{tg}$ - approaches zero at a single point within one SAW wavelength. Therefore, $n_{peak}$ occurs when carrier stripes are formed, we will discuss the stripe formation below. The black dashed line in Fig. 2d represents Equation VIII for $v = v_{sound}$, showing that $v$ does not exceed this ultimate carrier velocity in LiNbO$_3$ at any density. For $n < n_{peak}$, $J_{AE}$ saturates $v = 3,700\ m/s$, consistent with the nonlinear regime. Therefore, nonlinear acoustoelectric transport behavior plays a major role for $n < n_{peak}$ at high $I$ ($\Phi > V_{tg}$). While for $n > n_{peak}$, $J_{AE} \sim 1/n$, consistent with Equation VII in the linear regime ($\Phi < V_{tg}$).

Figures 3a and 3b show $J_{AE}/I$ (proportional to $\Gamma$, per Equation I) vs. $n$ for holes and electrons. For $n < n_{peak}$, $J_{AE}/I$ increases linearly with $n$ before peaking, with steeper slopes at lower as $J_{AE}/I \sim n/I$, deviating from Equation V. For $n > n_{peak}$, $J_{AE}/I$ decreases with increasing $n$, independent of $I$, as $J_{AE}/I \sim 1/n$ in agreement with Equation V. The transition in the behavior of $J_{AE}/I$ vs. $n$ agrees well with the transition from linear to nonlinear behavior occurring at $n_{peak}$. Figures 3c and 3d depict $J_{AE}/I$ vs. $I$ for holes and electrons at various $n$. At low $n$ (e.g. $n = 1 \times 10^{11}$ cm$^{-2}$), $J_{AE}/I$ follows a $I^{-1}$ dependence, particularly for $I > 5$ W/m. As $n$ increases, $J_{AE}/I$ rises, confirming $J_{AE}/I \sim n/I$. At high $n$ (e.g. $n = 4 \times 10^{11}$ cm$^{-2}$), $J_{AE}/I$ shows a weak dependence on $I$. Similar trends in $J_{AE}/I$ vs. $n$ and $I$ [49,52] have been observed in heterojunctions of 2DEGs on LiNbO$_3$ substrates with enhanced $K^2$, attributed to carrier velocity saturation as $v$ approaches $v_{sound}$ in the nonlinear regime. The observed $J_{AE}/I \sim \Gamma$ vs. $n$ and $I$ can be well explained by our simulations as shown in the Supporting Fig. 1c.

For 2DEGs in the nonlinear regime, the SAW potential $\Phi$ formed large density modulations that were best characterized by the bunching of electrons into stripes. Our measurements suggest that large modulations of electrons or holes are generated in our sample in the nonlinear regime, as depicted in the schematic of Figure 3e. These modulations create stripes of single carriers at $n_{peak}$, where $\Phi = V_{tg}\frac{t_{BN}^b}{t_{BN}^t}$ (with $\frac{t_{BN}^b}{t_{BN}^t}$ the ratio of the bottom and top BN thicknesses), marking the upper density limit for nonlinear effects. At $n_{peak}$, within half a wavelength, carriers accumulate, while in the other half, they are depleted, resulting in a spatially alternating carrier distribution. The presence of such periodic carrier modulations can influence electronic transport in graphene. For $n < n_{peak}$, at high $I$, the system is in the nonlinear regime, and non-compensated electron and hole stripes are formed. While at the CNP, equal amounts of electrons and holes are formed into stripes. For $n > n_{peak}$, the density modulations weaken, and the system transitions to the linear regime. Unlike 2DEGs, graphene uniquely enables this spatial control of both electrons and holes to form stripes at low densities and high $I$.

We characterize the electrical properties of graphene in the presence of high-intensity counter-propagating SAWs, which generate SSAWs to demonstrate our ability to control these density modulations. Figure 4a shows the dependence of $I_{AE}$ vs. $I_R$ ($I$ generated by the right IDT) and the RF phase $\phi_L$ applied to the left IDT, with $I_L = 5.91$ W/m ($I$ generated by the right IDT) and $\phi_R = 0$ (RF phase applied to the right IDT). When the counter-propagating SAWs are phase-matched and of equal intensity, SSAWs form beneath the device are generated, causing $I_{AE}$ to vanish. These parameters are unique for each set of electrodes. Figure 4b presents horizontal line cuts of Figure 4a at four different $I_R$, showing periodic phase dependence, indicating SSAW is obtained as a function of $I_R$ and $\phi_L$. By adjusting $I_R$, we optimize the intensity combination to symmetrize the minima and maxima of the phase dependence. Figure 4c maps the combinations of $I_L$ and $I_R$ that result in SSAWs. Figure 4d demonstrates the tunability of SSAWs by sweeping the phase offset of the counter-propagating SAWs, enabling multiple SSAWs conditions through variations in $\phi_L$ and $\phi_R$. To examine their impact on electronic transport, SSAWs are positioned between selected contacts. Figure 4e illustrates the effects of SSAWs on the four-probe resistance ($R_{4p}$) vs. $V_{tg}$, showing resistance peaks for $n \sim n_{peak}$. These peaks disappear when SAWs are turned off, confirming the impact of the carrier stripes formed by the SSAWs on the electronic transport. Figure 4f shows the dependence of the change in resistance $\Delta R_{4p} = R_{4p}$ (SAWs ON) $- R_{4p}$ (SAWs OFF) as a function of $V_{tg}$ and

$\phi_L$. The resistance peaks near $n_{peak}$ are tuned by $\phi_L$, as SSAWs formation is highly phase sensitive. These results confirm that SSAWs modulate carrier distribution, acting as a dynamic gate to control electronic transport.

**Discussions**

For SAWs to serve as a platform for quantum simulation, they must provide sufficient control and trapping potentials. Our findings suggest that nonlinear acoustoelectric effects play a key role in achieving this. In graphene, SAW-induced piezoelectric fields ($E_{SAW}$) generate dissipative currents that attenuate the wave, described by $\Gamma = \langle jE_{SAW}\rangle/I$, where the brackets denote averaging over one wavelength. At low $I$, $\Gamma \sim 1/n$ depending strongly on $n$, as shown in Equation V. In single-layer graphene, however, $\Gamma$ peaks when $n$ approaches the disorder-limited carrier density, as discussed in the Supporting Information. At high $I$, strong carrier modulations (Figure 3e) reduce $\Gamma$. At low $n$, carriers become trapped in the SAW potential, forming stripes that propagate at $v_{sound}$. The absorbed SAW energy $\langle jE_{SAW}\rangle$ saturates, leading to a transition from $\Gamma \sim 1/n$ in the linear regime to $\Gamma = \frac{\langle jE_{SAW}\rangle}{I} = \frac{env_{sound}^2}{\mu I} \sim \frac{n}{I}$ in the nonlinear regime, as confirmed in our measurements of $\frac{J_{AE}}{I} \sim \Gamma$ (Figures 3a-d). At high $n$, the carrier density becomes too large to bunch into stripes, causing $v$ to depend on $I$, remaining consistent with that observed at the linear regime.

Supporting Fig. 2a shows the predicted $J_{AE}$ vs. $n$ assuming different disorder potentials $\Delta V$, with lower disorder yielding larger $J_{AE}$. Supporting Fig. 2b shows that higher $\Delta V$ requires stronger $I$ to enter the nonlinear regime, as the disorder potential must be surpassed. Our experimental data matches simulations with $\Delta V < 0.1V$, consistent with the high mobility of ~ 100,000 cm²/V.s of our graphene devices. In the linear model, for a fixed $\Delta V$, increasing $I$ only enhances the magnitude of $J_{AE}$ and $n_{peak}$ remains unchanged. We should also note that the reduced IL due to the optimized IDT designs helps to achieve the nonlinear regime.

At low $I$ ($I < (1.5)^2 W/m$), $n_{peak}$ remains unchanged, consistent with the linear regime. For $I > (1.5)^2 W/m$, $n_{peak}$ increases linearly with $\sqrt{I} \sim \Phi$ (inset of Figure 2d), indicating that SAWs act as dynamic gates. Figure 3e shows the predicted SAW-modulated carrier density in graphene (assuming a low $\Delta V$) where $V_{tg}$ can be thought of as a uniform offset. Therefore, SAWs generate single carrier stripes when the amplitude of the SAW-induced carriers is equal to the carrier density generated by $V_{tg}$. Thus, $n_{peak}$ is the density at which single carriers form stripes and marks the threshold between the linear and nonlinear regime.

**Conclusions**

We investigated nonlinear acoustoelectric transport in BN-encapsulated graphene devices on LiNbO₃ substrates, demonstrating that high-intensity SAWs induce carrier modulation and stripe formation. The transition from linear ($\Gamma \sim 1/n$) to nonlinear ($\Gamma \sim n/I$) transport, along with the saturation of $J_{AE}$, and the dependence of $n_{peak} \sim \sqrt{I}$, provides direct evidence of SAW-driven periodic charge structures. Unlike 2DEGs, graphene supports periodic stripes of electrons, holes, or electron-hole mixtures, enriching carrier dynamics. The formation of these stripes is marked by $n_{peak}$, which corresponds to the threshold where the carrier modulation induced by SAWs matches the modulation imposed by $V_{tg}$. High mobility (~100,000 cm²/V·s) and low disorder potential ($\Delta V < 0.1V$) in our devices enabled clear observation of nonlinear transport. Optimized IDT designs further enhanced the efficiency of SAW-carrier interactions. Using counter-propagating SAWs, we generated SSAWs, creating tunable charge distributions. The suppression of acoustoelectric current ($I_{AE} = 0$) under SSAWs conditions and the emergence of resistance peaks ($R_{4p}$) near $n_{peak}$ confirm carrier stripe formation and its impact on transport. These SSAW-induced stripes can be dynamically tuned by varying the SAW phase and intensity, enabling precise control of charge modulation. The ability to manipulate carrier dynamics using SAWs enables applications in quantum simulation, artificial charge density waves, and acoustic lattices for anyon braiding. Our results establish SAWs as a powerful tool for exploring acoustically induced phenomena in atomically thin materials, offering opportunities to study and engineer time-dependent quantum systems and nematic phases.

**References**

1.  Haldane, F. D. M. 'Luttinger liquid theory' of one-dimensional quantum fluids. I. Properties of the Luttinger model and their extension to the general 1D interacting spinless Fermi gas. *J. Phys. C Solid State Phys.* **14**, 2585 (1981).

2.  Luttinger, J. M. An Exactly Soluble Model of a Many-Fermion System. *J. Math. Phys.* **4**, 1154–1162 (1963).

3.  Tomonaga, S. Remarks on Bloch's Method of Sound Waves applied to Many-Fermion Problems. *Prog. Theor. Phys.* **5**, 544–569 (1950).

4.  Devereaux, T. P. & Kivelson, S. A. The significance of 'stripes' in the physics of the cuprates, the Hubbard model, and other highly correlated electronic systems. ArXiv.2501.15709 (2025).

5.  Emery, V. J., Kivelson, S. A. & Tranquada, J. M. Stripe phases in high-temperature superconductors. *Proc. Natl. Acad. Sci.* **96**, 8814–8817 (1999).

6.  Zaanen, J. & Gunnarsson, O. Charged magnetic domain lines and the magnetism of high-$T_c$ oxides. *Phys. Rev. B* **40**, 7391–7394 (1989).

7.  Lu, X., Chen, F., Zhu, W., Sheng, D. N. & Gong, S.-S. Emergent Superconductivity and Competing Charge Orders in Hole-Doped Square-Lattice t-J Model. *Phys. Rev. Lett.* **132**, 066002 (2024).

8.  Xi, X. *et al.* Strongly enhanced charge-density-wave order in monolayer $NbSe_2$. *Nat. Nanotechnol.* **10**, 765–769 (2015).

9.  Burk, B., Thomson, R. E., Clarke, J. & Zettl, A. Surface and Bulk Charge Density Wave Structure in 1 T-$TaS_2$. *Science* **257**, 362–364 (1992).

10. Grüner, G. The dynamics of charge-density waves. *Rev. Mod. Phys.* **60**, 1129–1181 (1988).

11. McMillan, W. L. Theory of discommensurations and the commensurate-incommensurate charge-density-wave phase transition. *Phys. Rev. B* **14**, 1496–1502 (1976).

12. Fradkin, E., Kivelson, S. A. & Tranquada, J. M. Colloquium: Theory of intertwined orders in high temperature superconductors. *Rev. Mod. Phys.* **87**, 457–482 (2015).

13. Tsui, D. C., Stormer, H. L. & Gossard, A. C. Two-Dimensional Magnetotransport in the Extreme Quantum Limit. *Phys. Rev. Lett.* **48**, 1559–1562 (1982).

14. You, Y., Cho, G. Y. & Fradkin, E. Theory of Nematic Fractional Quantum Hall States. *Phys. Rev. X* **4**, 041050 (2014).

15. Maciejko, J., Hsu, B., Kivelson, S. A., Park, Y. & Sondhi, S. L. Field theory of the quantum Hall nematic transition. *Phys. Rev. B* **88**, 125137 (2013).

16. Lilly, M. P., Cooper, K. B., Eisenstein, J. P., Pfeiffer, L. N. & West, K. W. Evidence for an Anisotropic State of Two-Dimensional Electrons in High Landau Levels. *Phys. Rev. Lett.* **82**, 394–397 (1999).

17. Mulligan, M., Nayak, C. & Kachru, S. Isotropic to anisotropic transition in a fractional quantum Hall state. *Phys. Rev. B* **82**, 085102 (2010).

18. Fradkin, E., Kivelson, S. A., Lawler, M. J., Eisenstein, J. P. & Mackenzie, A. P. Nematic Fermi Fluids in Condensed Matter Physics. *Annu. Rev. Condens. Matter Phys.* **1**, 153–178 (2010).

19. Jin, C. *et al.* Stripe phases in $WSe_2/WS_2$ moiré superlattices. *Nat. Mater.* **20**, 940–944 (2021).

20. Rubio-Verdú, C. *et al.* Moiré nematic phase in twisted double bilayer graphene. *Nat. Phys.* **18**, 196–202 (2022).

21. Liu, S.-B. *et al.* Nematic Ising superconductivity with hidden magnetism in few-layer 6R-$TaS_2$. *Nat. Commun.* **15**, 7569 (2024).

22. Cao, L. *et al.* Directly visualizing nematic superconductivity driven by the pair density wave in $NbSe_2$. *Nat. Commun.* **15**, 7234 (2024).

23. Laroche, D., Gervais, G., Lilly, M. P. & Reno, J. L. 1D-1D Coulomb Drag Signature of a Luttinger Liquid. *Science* **343**, 631–634 (2014).

24. Yu, G. *et al.* Evidence for two dimensional anisotropic Luttinger liquids at millikelvin temperatures. *Nat. Commun.* **14**, 7025 (2023).

25. Wu, Y.-M., Murthy, C. & Kivelson, S. A. Possible Sliding Regimes in Twisted Bilayer WTe$_2$. *Phys. Rev. Lett.* **133**, 246501 (2024).

26. Wang, P. *et al.* One-dimensional Luttinger liquids in a two-dimensional moiré lattice. *Nature* **605**, 57–62 (2022).

27. Gupta, S., Yu, H. & Yakobson, B. I. Designing 1D correlated-electron states by non-Euclidean topography of 2D monolayers. *Nat. Commun.* **13**, 3103 (2022).

28. Liu, J. *et al.* Spin-Triplet Excitonic Insulator in the Ultra-Quantum Limit of HfTe$_5$. ArXiv.2501.12572 (2025).

29. Wu, W. *et al.* Topological Lifshitz transition and one-dimensional Weyl mode in HfTe$_5$. *Nat. Mater.* **22**, 84–91 (2023).

30. Du, X. *et al.* Crossed Luttinger liquid hidden in a quasi-two-dimensional material. *Nat. Phys.* **19**, 40–45 (2023).

31. Byrnes, T., Recher, P., Kim, N. Y., Utsunomiya, S. & Yamamoto, Y. Quantum Simulator for the Hubbard Model with Long-Range Coulomb Interactions Using Surface Acoustic Waves. *Phys. Rev. Lett.* **99**, 016405 (2007).

32. Schuetz, M. J. A. *et al.* Acoustic Traps and Lattices for Electrons in Semiconductors. *Phys. Rev. X* **7**, 041019 (2017).

33. Wu, M. *et al.* Probing Quantum Phases in Ultra-High-Mobility Two-Dimensional Electron Systems Using Surface Acoustic Waves. *Phys. Rev. Lett.* **132**, 076501 (2024).

34. Sogawa, T., Sanada, H., Gotoh, H., Yamaguchi, H. & Santos, P. V. Dynamic control of photoluminescence polarization properties in GaAs/AlAs quantum wells by surface acoustic waves. *Phys. Rev. B* **86**, 035311 (2012).

35. Wixforth, A. *et al.* Surface acoustic waves on GaAs/Al$_x$Ga$_{1-x}$As heterostructures. *Phys. Rev. B* **40**, 7874–7887 (1989).

36. Esslinger, A. *et al.* Acoustoelectric study of localized states in the quantized Hall effect. *Solid State Commun.* **84**, 939–942 (1992).

37. Mou, Y. *et al.* Gate-Tunable Quantum Acoustoelectric Transport in Graphene. *Nano Lett.* **24**, 4625–4632 (2024).

38. Fang, Y. *et al.* Quantum Oscillations in Graphene Using Surface Acoustic Wave Resonators. *Phys. Rev. Lett.* **130**, 246201 (2023).

39. Nichols, D. M. *et al.* Charge pumping in h-BN-encapsulated graphene driven by surface acoustic waves. *J. Appl. Phys.* **136**, 024302 (2024).

40. Zhao, P. *et al.* Acoustically Induced Giant Synthetic Hall Voltages in Graphene. *Phys. Rev. Lett.* **128**, 256601 (2022).

41. Preciado, E. *et al.* Scalable fabrication of a hybrid field-effect and acousto-electric device by direct growth of monolayer $MoS_2$/$LiNbO_3$. *Nat. Commun.* **6**, 8593 (2015).

42. Yokoi, M. *et al.* Negative resistance state in superconducting $NbSe_2$ induced by surface acoustic waves. *Sci. Adv.* **6**, eaba1377 (2020).

43. Peng, R. *et al.* Long-range transport of 2D excitons with acoustic waves. *Nat. Commun.* **13**, 1334 (2022).

44. Hernández-Mínguez, A., Liou, Y.-T. & Santos, P. V. Interaction of surface acoustic waves with electronic excitations in graphene. *J. Phys. Appl. Phys.* **51**, 383001 (2018).

45. Willett, R. L. & Pfeiffer, L. N. Composite fermions examined with surface acoustic waves. *Surf. Sci.* **361–362**, 38–41 (1996).

46. Utko, P., Hansen, J. B., Lindelof, P. E., Sørensen, C. B. & Gloos, K. Single-Electron Transport Driven by Surface Acoustic Waves: Moving Quantum Dots Versus Short Barriers. *J. Low Temp. Phys.* **146**, 607–627 (2007).

47. Shilton, J. M. *et al.* High-frequency single-electron transport in a quasi-one-dimensional GaAs channel induced by surface acoustic waves. *J. Phys. Condens. Matter* **8**, L531 (1996).

48. Rotter, M., Wixforth, A., Ruile, W., Bernklau, D. & Riechert, H. Giant acoustoelectric effect in GaAs/$LiNbO_3$ hybrids. *Appl. Phys. Lett.* **73**, 2128–2130 (1998).

49. Rotter, M. *et al.* Nonlinear acoustoelectric interactions in GaAs/LiNbO$_3$ structures. *Appl. Phys. Lett.* **75**, 965–967 (1999).

50. White, D. L. Amplification of Ultrasonic Waves in Piezoelectric Semiconductors. *J. Appl. Phys.* **33**, 2547–2554 (1962).

51. Poole, T. & Nash, G. R. Acoustoelectric Current in Graphene Nanoribbons. *Sci. Rep.* **7**, 1767 (2017).

52. Rotter, M., Kalameitsev, A. V., Govorov, A. O., Ruile, W. & Wixforth, A. Charge Conveyance and Nonlinear Acoustoelectric Phenomena for Intense Surface Acoustic Waves on a Semiconductor Quantum Well. *Phys. Rev. Lett.* **82**, 2171–2174 (1999).

**Acknowledgments**

L.A.J. acknowledges the support from the NSF-CAREER (DMR 2146567). T.M. acknowledges support from the Eddleman Quantum Institute (EQI) through the EQI graduate research fellowship. K.S. acknowledges support from EQI and the Undergraduate Research Opportunities program through the QUROP fellowship. The work of D.T. and T.S. was supported by the U.S. Department of Energy, Office of Science, Office of Basic Energy Sciences under Award Number DE-SC0025568. This work was performed, in part, at the Center for Integrated Nanotechnologies, an Office of Science user facility operated for the U.S. Department of Energy (DOE) Office of Science. SNL is a multimission laboratory managed and operated by National Technology and Engineering Solutions of Sandia LLC, a wholly owned subsidiary of Honeywell International Inc. for the U.S. DOE National Nuclear Security Administration under contract number DE-NA0003525. This paper describes objective technical results and analysis. Any subjective views or opinions that might be expressed in the paper do not necessarily represent the views of the U.S. DOE or the United States government. We thank Javier Sanchez-Yamagishi for discussions during the initial experiments. We thank Qiyin Lin for assistance with the metal evaporation using the Angstrom e-beam evaporator at the UC Irvine Materials Research Institute (IMRI); Matthew Law and Geemin Kim for helping us using their thermal metal evaporator; and SungWoo Nam for helping with their e-beam evaporator.

**Competing interests**

The authors declare no competing interests.

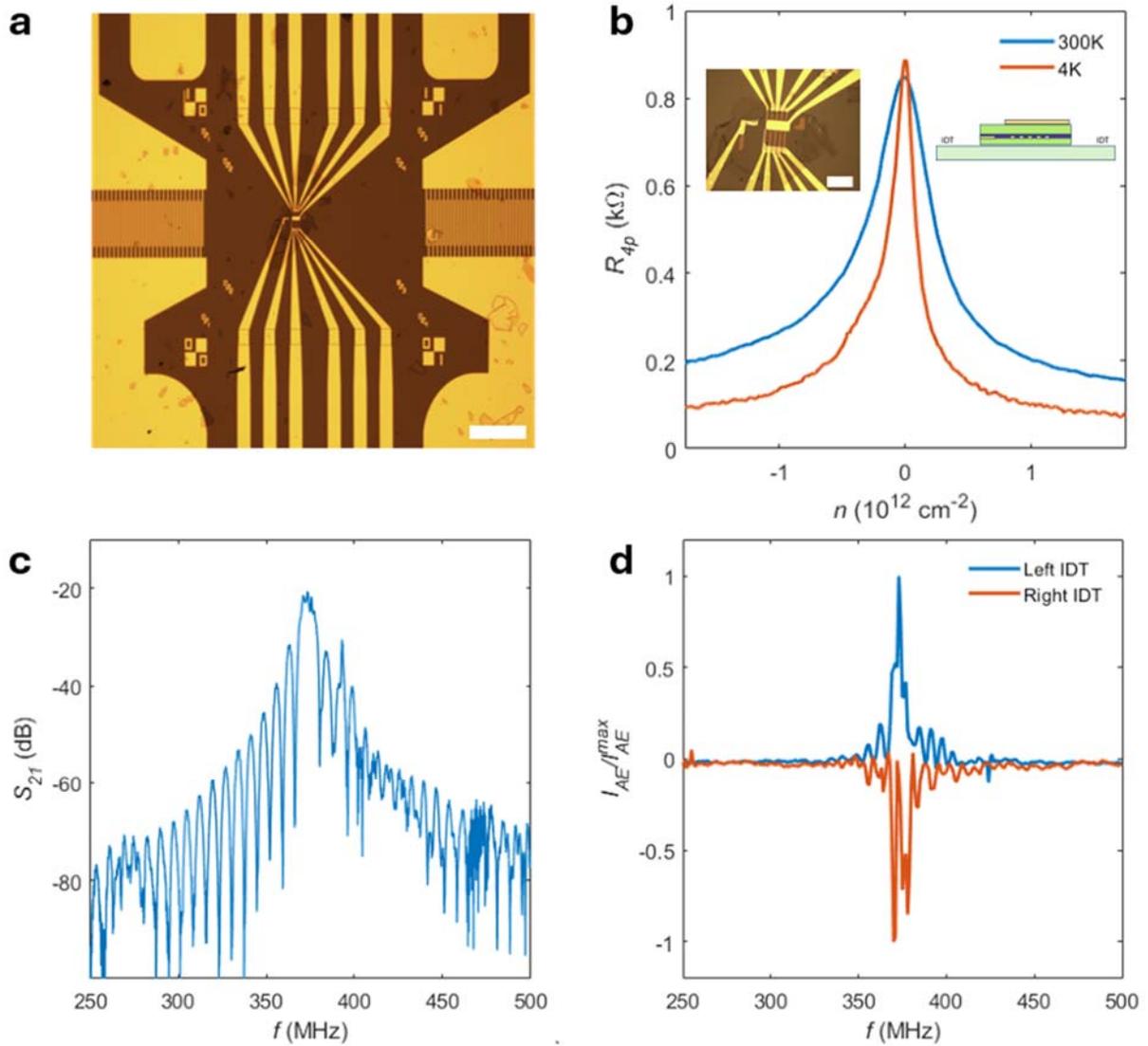

**Figure 1. High-quality graphene device on a LiNbO₃ substrate. a** Optical image of a boron nitride (BN)-encapsulated graphene device (Device #1) positioned between two interdigitated transducers (IDTs) fabricated on a LiNbO₃ substrate. Scale bar: 100 $\mu$m. **b** Four-probe resistance ($R_{4p}$) vs. carrier density ($n$) induced by the top gate, measured at $T = 300$K and 4K. Left inset: Optical image of the device. Scale bar: 20 $\mu$m. Right inset: Schematic of the heterostructure cross section. The top and bottom BN thicknesses are 25nm and 15nm, respectively, with the heterostructure placed on pre-patterned Cr/PdAu bottom contacts (1/10nm). **c** Transmission spectrum ($S_{21}$) measured at $T = 300$K, characterizing the insertion-loss ($IL$) of the IDTs, measured with the device positioned between the IDTs. **d** Normalized acoustoelectric current ($I_{AE}/I_{AE}^{max}$) vs. frequency ($f$), measured at $T = 300$K. A tunable frequency generator with 15 dBm of output power was applied to either the left or right IDTs. A sharp increase in $I_{AE}$ is observed near the IDT's fundamental resonance frequency $f \sim 375$ MHz, consistent with the transmission spectrum in **c**.

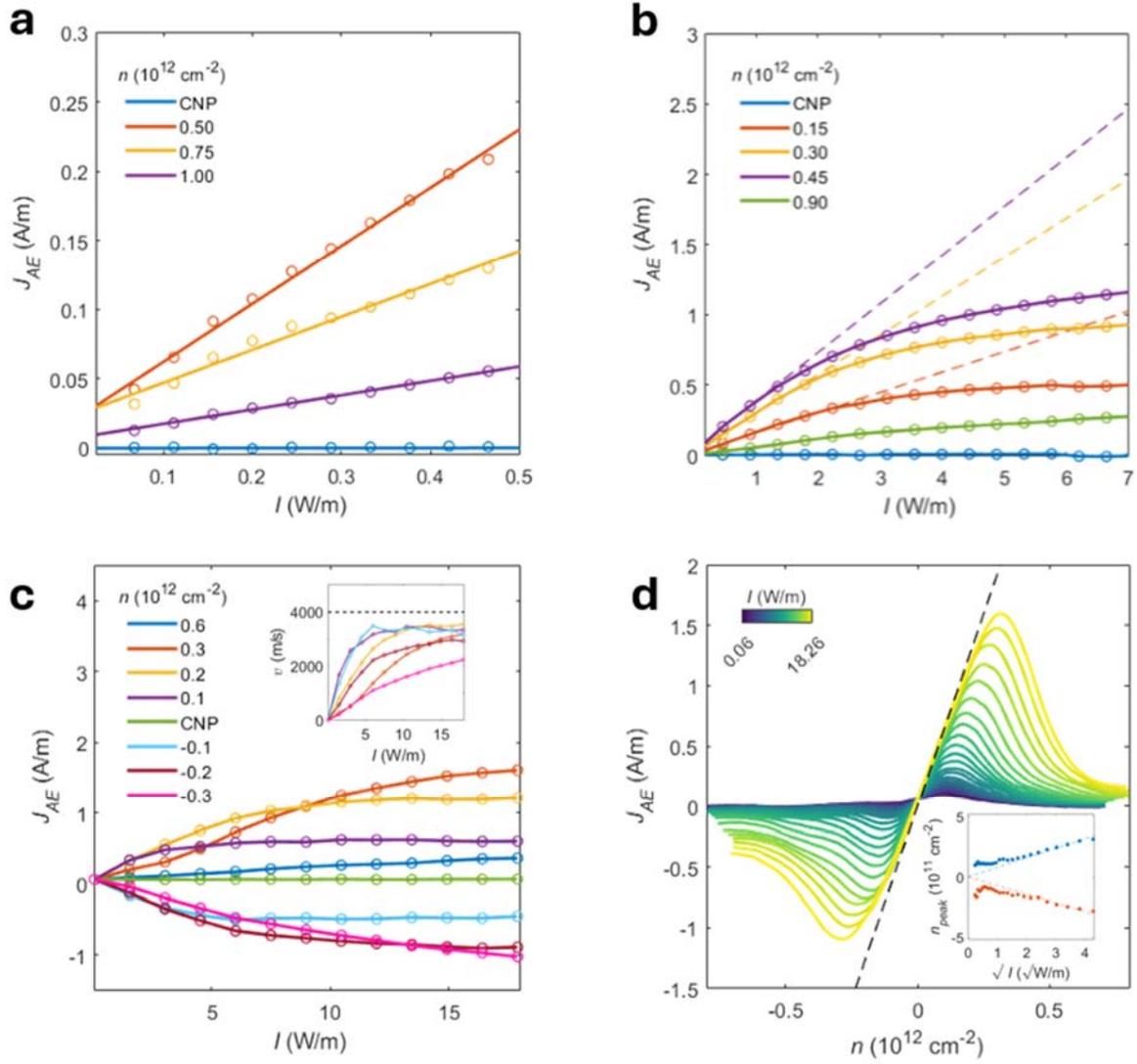

**Figure 2. Intensity ($I$) and carrier density ($n$) dependence of the acoustoelectric current ($J_{AE}$). a** $J_{AE}$ vs. $I$ measured at several electron densities ($n$) for $I$ up to 0.5 W/m. **b** $J_{AE}$ vs. $I$ measured at various $n$ for $I$ up to 7W/m. Dashed lines represent linear fits for $I$ up to 0.5 W/m. Both **a** and **b** were measured at $T$ = 300K. **c** $J_{AE}$ vs. $I$ measured at various $n$ for up to 18 W/m. Inset: extracted carrier velocity ($v$) vs. $I$ for the same densities (excluding $n$ = 0), color-coded as in the legend. The horizontal black dashed line corresponds to $v_{sound}$ = 3,980 m/s. **d** $J_{AE}$ vs. $n$ measured at various intensities. The black dashed line corresponds to $J_{AE} = env_{sound}$, where $e$ is the electron charge. Inset: extracted $n_{peak}$, the carrier density at which $J_{AE}$ peaks (for electrons and holes), plotted against $\sqrt{I}$. Dashed lines represent linear fits. **c** and **d** were measured at $T$ = 4K. All the measurements shown in this figure were performed using the right IDT.

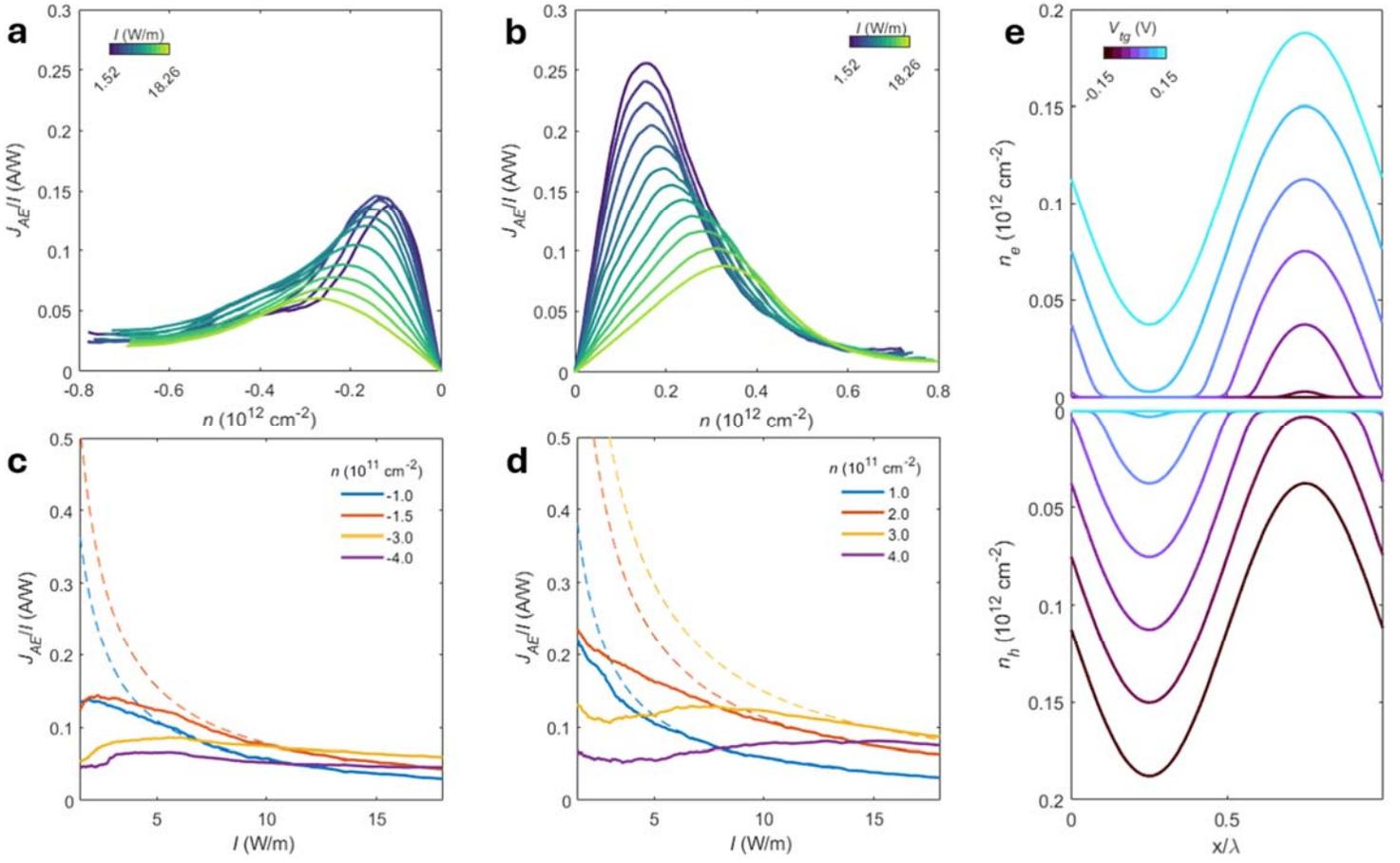

**Figure 3. Nonlinear attenuation and formation of carrier stripes. a** $J_{AE}/I$ vs. $n$ measured at various SAW intensities on the hole side. **b.** Same as **a** but measured on the electron side. **c** $J_{AE}/I$ vs. $I$ measured at various hole densities. The dashed lines represent fits to $1/I$. **d** Same as **c** but measured for various electron densities. All measurements shown in **a-d** were conducted at $T = 4$K. **e** Calculated spatial modulation of the electron (top) and hole (bottom) densities vs. $x'/\lambda$, where $x'$ is a representative distance in the sample which is less than $\lambda = 10$ $\mu$m, as a function of $V_{tg}$ with a constant SAW amplitude $\Phi = 0.1$V and disorder potential $\Delta V = 0.01$V.

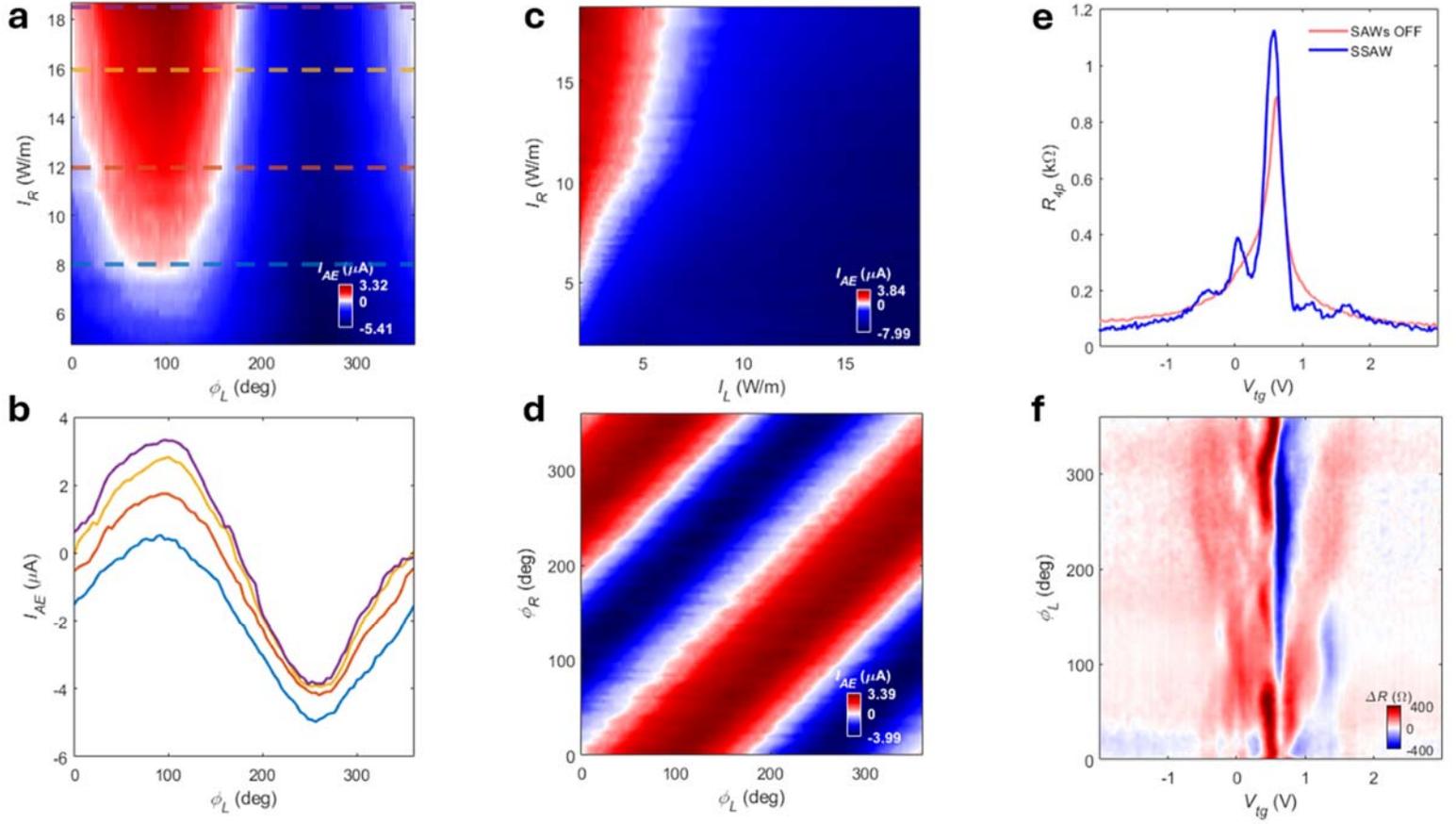

**Figure 4. Formation and utilization of standing surface acoustic waves. a** $I_{AE}$ vs. $I_R$ (SAW intensity generated by the right IDT) and phase from the left IDT ($\phi_L$). With $I_L$ (SAW intensity generated by the left IDT) = 5.91 W/m. **b** Line cuts of $I_{AE}$ vs. $\phi_L$, corresponding to the dashed lines in **a**. **c** $I_{AE}$ vs. $I_L$ and $I_R$, with a fixed phase difference of 4 degrees. $I_{ae} = 0$ reveal combinations of $I_L$ and $I_R$ that achieve standing SAWs (SSAWs) conditions. **d** $I_{AE}$ vs. $\phi_L$ and phase from the right IDT ($\phi_R$), with fixed $I_L$ = 5.91 W/m and $I_R$ = 18.69 W/m. **e** Four probe resistance ($R_{4p}$) vs. $V_{tg}$ with and without SSAWs. Counter-propagating SAW intensities are using the same fixed combination as **c**, and the same phase offset as **d**. **f** Change in $R_{4p}$, $\Delta R_{4p}$ ($R_{4p}$ (SAWs ON) - $R_{4p}$ (SAWs OFF)) vs. $V_{tg}$ and $\phi_L$ in the presence of counter-propagating SAWs, (using the $I_L$ and $I_R$ as defined in **c**). For **d-f**, $\phi_R = 0$.